\documentclass[11pt]{article}
\usepackage[margin=1in]{geometry}

\usepackage{amsmath, amssymb, amsthm}
\usepackage{proof}
\usepackage{stmaryrd}
\usepackage{url} 

\newcommand{\subsymb}{\mbox{\tt <\!:}}
\newcommand{\subbndsymb}{\mbox{\tt <\!:\!}_b}
\newcommand{\sub}[2]{#1\, \subsymb\, #2}
\newcommand{\subbnd}[2]{#1\, \subbndsymb\, #2}
\newcommand{\suptype}{\mbox{\tt Top}}

\newcommand{\satop}{\mbox{\sc SA-Top}}
\newcommand{\sareflvar}{\mbox{\sc SA-Refl-TVar}}
\newcommand{\satransvar}{\mbox{\sc SA-Trans-TVar}}
\newcommand{\saarrow}{\mbox{\sc SA-Arrow}}
\newcommand{\saall}{\mbox{\sc SA-All}}
\newcommand{\sahyp}{\mbox{\sc SA-Hyp}}
\newcommand{\satrvar}{\mbox{\sc SA-Tr-TVar}}

\newtheorem{proposition}{Prop}

\title{About a Proof Pearl: A Purported Solution to \\
  a {\sc PoplMark} Challenge Problem that Is Not One}
\author{Gopalan Nadathur}

\begin{document}
\maketitle

\begin{abstract}
\noindent The {\sc PoplMark} Challenge comprises a set of problems
intended to measure the strength of reasoning systems in the
realm of mechanizing programming language meta-theory at the time the
challenge was enunciated.
Included in the collection is the exercise of demonstrating 
transitivity of subtyping for a specific algorithmic formulation
of subtyping for an extension of System F.
The challenge represented by this problem derives 
from the fact that, for the given formulation, subtyping must
be proved simultaneously with another property called narrowing.
In a paper published as a proof pearl, Brigitte Pientka claimed to
have presented a solution to the problem in which ``the full power of
parametric and higher-order judgments'' is exploited to ``get the
narrowing lemma for free.''
We show this claim to be inaccurate. 
In particular, we show that the simplification is in substantial part
the result of changing the formulation of the subtyping relation in a
way that modifies the challenge rather than the outcome of the manner in 
which the argument is mechanized.
\end{abstract}

\noindent The {\sc PoplMark} Challenge~\cite{aydemir05tphols}
identified a collection of benchmarks for measuring the status of
reasoning frameworks at the time that it was presented from the
perspective of mechanizing the meta-theory of programming languages.
The problems in this collection are oriented around the typed
$\lambda$-calculus with second-order polymorphism known as System F
extended with subtyping and records.
The formalization task that is of specific interest here is
what is referred to in the mentioned paper as the transitivity of
algorithmic subtyping.
This problem, which was identified as \emph{Challenge 1A}, was
considered worthy of inclusion in the benchmarks because the
proof of transitivity requires an intricate inductive argument in
which an auxiliary lemma called the \emph{narrowing lemma} must be
established simultaneously.
In~\cite{pientka07tphols}, Brigitte Pientka claimed to have presented
a solution in the Twelf system that ``plays to the strengths of the
logical framework LF'' and that as a benefit gets the ``tedious
narrowing lemma, which must normally be proved separately, for free.''
Unfortunately, what is really presented does not constitute a solution
to Challenge 1A.
Specifically, the ``simplification'' in the proof results not from the
choice of encoding but from changing the formalization of subtyping to
a form that comes close to including narrowing in a typing rule.

We provide substance to the above observations in the rest of this
note.
We first recall the original formalization task and then explain our
comments about the ``solution'' presented in~\cite{pientka07tphols}.

\paragraph{The formalization task}
Challenge 1A relates solely to the type language that includes
subtyping but does not encompass record types.
The two syntactic entities that are relevant to describing the
subtyping relation in this context are types and type environments.
Expressions in these categories are given by the following grammar
rules:
\[
  \begin{array}{rrcl}
    \mbox{\it Types} &\quad T & ::= &
           X\ |\ \suptype\ |\ T \rightarrow T\ |\ \forall \sub{X}{T}.T\\[3pt] 
    \mbox{\it Type Environments} &\quad \Gamma & ::= &
           \cdot\ |\ \Gamma, \subbnd{X}{T}
  \end{array}
\]
The token $X$ represents variables in these rules.
A type of the form $\forall \sub{X}{T_1}.T_2$ represents a binding of
$X$ over the type $T_2$.
Type environments bind variables that occur in types within their
scope and they identify subtyping constraints for the variables so
bound.
Variables bound by an environment are assumed to be named uniquely
and the subtyping rules implicitly permit the renaming of bound
variables in types to ensure that this property holds.
The presentation in~\cite{aydemir05tphols} uses the same symbol for
both $\subsymb$ and $\subbndsymb$.
We avoid this overloading here to highlight issues that arise 
from the confusion of the two in~\cite{pientka07tphols}.

\begin{figure}[tbhp]
\begin{center}
  \begin{tabular}{cc}
    \infer[\satop]
          {\Gamma \vdash \sub{S}{\suptype}}
          {}

    \quad
    &
    \quad

    \infer[\sareflvar]
          {\Gamma \vdash \sub{X}{X}}
          {}

  \end{tabular}

  \vspace{10pt}
  
  \begin{tabular}{cc}
    \infer[\satransvar]
          {\Gamma \vdash \sub{X}{T}}
          {\subbnd{X}{U} \in \Gamma \qquad \Gamma \vdash \sub{U}{T}}

    \quad
    &
    \quad
    \infer[\saarrow]
          {\Gamma \vdash \sub{S_1 \rightarrow S_2}
                             {T_1 \rightarrow T_2}}
          {\Gamma \vdash \sub{T_1}{S_1}
            \qquad
           \Gamma \vdash \sub{S_2}{T_2}}
  \end{tabular}

  \vspace{10pt}

  \begin{tabular}{c}
    \infer[\saall]
          {\Gamma \vdash \sub{\forall \sub{X}{S_1}.S_2}
                             {\forall \sub{X}{T_1}.T_2}}
          {\Gamma \vdash \sub{T_1}{S_1} \qquad
            \Gamma,\subbnd{X}{T_1} \vdash \sub{S_2}{T_2}}
  \end{tabular}
\end{center}

\caption{Rules Defining the Subtyping Relation}
\label{fig:subtyping}
\end{figure}

The subtyping relation that is of interest is written as $\Gamma
\vdash \sub{S}{T}$, to be read as ``$S$ is a subtype of $T$ relative
to the type environment $\Gamma$.''
This relation is determined by the rules in Figure~\ref{fig:subtyping}.
The $\satop$ rule has the requirement that $S$ must be
well-scoped with respect to $\Gamma$, i.e. all its free
variables must be bound in $\Gamma$.
Similarly, the $\sareflvar$ rule has the proviso that the variable $X$
must be bound in $\Gamma$.\footnote{It can be shown that these
  assumptions suffice to ensure that $\Gamma \vdash \sub{U}{V}$ is
  derivable only for types $U$ and $V$ that are well-scoped. It may be
  desirable to impose well-scoping requirements on $\Gamma$ as well,
  in which case the order in which bindings appear in it will be
  important. We discuss this observation further later in the note.}
This collection of rules has an algorithmic flavor in that it
determines a process for checking a purported subtyping relation that
is syntax-directed.
The form of the $\satransvar$ rule is crucial to this property.
Note that because of the form of this rule it is not automatically the
case that $\Gamma \vdash \sub{X}{U}$ is derivable if $\subbnd{X}{U}$
appears in $\Gamma$.
However, this turns out to be the case if $U$ is well-scoped with
respect to $\Gamma$ because then $\Gamma \vdash \sub{U}{U}$ is derivable.

Challenge 1A consists of providing a mechanized proof of the fact that
the subtyping relation defined by the rules in
Figure~\ref{fig:subtyping} is transitive. 
Specifically, a proof is to be provided for the following
proposition within a chosen system for formal reasoning:

\begin{proposition}\label{prop:trans}
If $\Gamma \vdash \sub{S}{Q}$ and $\Gamma \vdash \sub{Q}{T}$ are
derivable then so is $\Gamma \vdash \sub{S}{T}$.
\end{proposition}

The difficulty in showing this property, and the reason why it is
interesting as a challenge problem, is that it must be proved
simultaneously with the following property that is called narrowing:

\begin{proposition}\label{prop:narrowing}
If $\Gamma_1, \subbnd{X}{Q}, \Gamma_2 \vdash \sub{M}{N}$ and $\Gamma_1
\vdash \sub{P}{Q}$ are derivable then 
$\Gamma_1, \subbnd{X}{P}, \Gamma_2 \vdash \sub{M}{N}$ is also derivable.
\end{proposition}

In the standard informal proof, the properties are established by an
induction on the structure of $Q$, with the first property being shown
first and then used in showing the second.
The proof of the first property requires an additional induction on the
derivation of $\Gamma \vdash \sub{S}{Q}$ and that of the second
property similarly requires an additional induction on the derivation
of $\Gamma_1, \subbnd{X}{Q}, \Gamma_2 \vdash \sub{M}{N}$.

While the description of the challenge allows for small changes to the
subtyping rules, such as those needed to accommodate a higher-order
abstract syntax style of formalization, it places a requirement that
these such changes should result in an ``obviously equivalent''
system.
As an example, mention is made in~\cite{aydemir05tphols} of a
declarative presentation of subtyping that explicitly include rules
for reflexivity and transitivity.
Such a presentation obviates a proof of transitivity of subtyping and
thus constitutes another ``nice example how the original problem
specification influences the proofs about it'' in the sense hinted at
at the end of Section~2 in~\cite{pientka07tphols}. 
However, it also illustrates fairly starkly how changing the ``problem
specification'' in such ways is not acceptable because it undermines
the intended challenge.

\paragraph{The issues with the purported higher-order solution}

The claimed solution to the challenge in~\cite{pientka07tphols} uses
an encoding of the subtyping rules in LF that is based on a
higher-order abstract syntax style.
However, it is unnecessary to get into the details of this encoding to
illuminate the difficulty with the claims therein.
For simplicity of presentation, we will adhere to a presentation style
that is close to that of the original description of the challenge.
We note that this style is also used at the outset
in~\cite{pientka07tphols}. 

The presentation of subtyping in~\cite{pientka07tphols} begins by
confusing subtyping in bindings in type environments with subtyping
judgements. 
In the description of the challenge in this note, we have highlighted
the distinction between the two by deliberately using the symbol
$\subbndsymb$ for one and $\subsymb$ for the other.
In keeping with the presentation in~\cite{aydemir05tphols}, the symbol
$\subsymb$ is used in~\cite{pientka07tphols} for
both.\footnote{The symbol that is actually used
  in~\cite{pientka07tphols} is $\leq$ but we will assume it to be 
  $\subsymb$ here to ensure consistency with the earlier discussion.}
However, unlike in~\cite{aydemir05tphols} where a distinction is
maintained despite the overloading of notation, the two notions are
conflated in~\cite{pientka07tphols} by the inclusion of the following
rule:
\[
   \infer[\sahyp]
         {\Gamma \vdash \sub{X}{T}}
         {\sub{X}{T} \in \Gamma}
\]
A further distinction between the subtyping rules in the original
presentation and the one in~\cite{pientka07tphols} is that the latter
builds in transitivity with respect to variable subtyping directly
by replacing the $\satransvar$ rule with the following:
\[
  \infer[\satrvar]
        {\Gamma \vdash \sub{X}{T}}
        {\Gamma \vdash \sub{X}{U} \qquad \Gamma \vdash \sub{U}{T}}
\]

The changes to the formulation of subtyping that are described above
are substantial and do not satisfy the criterion of obvious
equivalence required by the challenge.
In particular, they considerably simplify the informal proof of
transitivity of subtyping.
Unlike with the original presentation of subtyping, the narrowing
lemma can be proved \emph{independently} of the transitivity property.
Further, Prop~\ref{prop:trans}, the transitivity property, can be
proved by an induction only on $Q$; the additional induction on the
derivation of $\Gamma \vdash \sub{S}{Q}$ is not needed.\footnote{The
  development in~\cite{pientka07tphols} continues to use an induction
  on both but the second induction can be avoided.}

In the above discussion, we have assumed that the narrowing lemma
still needs to be proved.
The claim in~\cite{pientka07tphols} is that this lemma is obtained for
free.
A closer look at the development shows that this ``simplification'' is
actually the result of assuming the validity of a further rule related
to subtyping that has the following form:
\[
\infer{\Gamma_1,\sub{X}{U},\Gamma_2 \vdash \sub{M}{N}}
      {\Gamma_1,\sub{X}{U}, \Gamma_2 \vdash \sub{X}{V} \qquad
       \Gamma_1, \sub{X}{V},\Gamma_2 \vdash \sub{M}{N}}
\]
Observe that the sensibility of this rule depends on the conflation of 
subtyping in bindings in type environments with the subtyping
judgement.
While the narrowing lemma is an easy consequence of this rule in a
situation where the $\satrvar$ rule is also present, the assumption of
such a rule represents a significant deviation from the formulation of
subtyping in Challenge 1A. 
We note that this rule can be shown to be admissible in the typing
calculus of interest, but doing this requires an argument that is
as complicated as the direct proof of narrowing.
This observation raises in a different way the question of 
what it means to get something for free.

\paragraph{A comment on ordering of bindings in type environments}

Typing calculi often attribute a significance to the order of bindings
in contexts or type environments.
In the discussions here, this sensitivity is manifest specifically in
the statement of Prop~\ref{prop:narrowing}.
While order is important for well-scoping of type environments, the
proofs of transitivity of subtyping and narrowing are not dependent on
order. 
Moreover, ignoring order has the effect of simplifying the
formalization of the arguments.
It is possible to realize such an effect in a system like
Abella~\cite{baelde14jfr}, which treats contexts by default as
multisets, by not imposing a well-scoping constraint initially on type 
environments. 
Once the desired properties have been proved when type environments
have this less constrained form, it is easy to show that they must
also hold when type environments are required to be well-scoped.

\section*{Acknowledgements}

The observations in this note were made in the course of conducting
research supported by the National Science Foundation under Grant 
No. CCF-1617771. 
Any opinions, findings, and conclusions or recommendations expressed
in this material are those of the authors and do not necessarily
reflect the views of the National Science Foundation.

\bibliographystyle{abbrv}
\bibliography{../references/master}
\end{document}